\documentstyle[ltwol,epsfig]{article}

\def\be{\begin{equation}}
\def\ee{\end{equation}}
\def\bea{\begin{eqnarray}}
\def\eea{\end{eqnarray}}

\begin{document}

\title {\null\vspace*{-.0cm}\hfill {\small nucl-th/0105063} \\ \vskip
0.8cm
RELATIVISTIC GENERALIZATION OF THE POST-PRIOR EQUIVALENCE FOR
REACTION OF COMPOSITE PARTICLES}

\author{CHEUK-YIN WONG$^1$ and HORACE W. CRATER$^2$}

\address{$^1$Physics Division, Oak Ridge National Laboratory, Oak
Ridge, Tennessee 37831\\ $^2$Department of Physics, University
of Tennessee Space Institute, Tullahoma, Tennessee 37388\\
E-mail: wongc@ornl.gov}


\twocolumn[ \maketitle\abstracts{ In the non-relativistic description
of the reaction of composite particles, the reaction matrix is
independent of the choice of post or prior forms for the interaction.
We generalize this post-prior equivalence to the relativistic reaction
of composite particles by using Dirac's constraint dynamics to
describe the bound states and the reaction process.}]

\section{Introduction}

In the non-relativistic description of composite particle reactions,
it is well known that the reaction matrix element is independent of
the choice of post or prior forms for the interaction \cite{Sch68}.
Specifically, for the reaction of composite particles $A$, $B$, $C$ and
$D$ with constituents 1, 2, 3, and 4,
\begin{equation}  \label{eq:1} A(12)+B(34)\rightarrow C(14)+D(32),
\end{equation}
the prior reaction matrix element between the initial state
$|i\rangle$ and the final state $|f\rangle$, $\langle f
|V_{13}+V_{14}+V_{23}+V_{24}|i\rangle$, is equal to the post reaction
matrix element, $\langle f |V_{13}+V_{12}+V_{43}+V_{42}|i\rangle$,
where $V_{ij}$ is the interaction between constituents $i$ and $j$.
This equality follows if and only if the interaction which induces the
reaction is the same as the interaction which generates the composite
bound states, and the reaction matrix element is evaluated using these
bound state wave functions.  The post-prior equivalence guarantees the
uniqueness of the reaction cross section in the Born approximation.

Recently there has been much interest in studying the reaction of
composite particles at relativistic energies in nuclear and particle
physics (see Refs.\ [2,3] and references cited therein).  It is
clearly of interest to derive a relativistic generalization of the
post-prior theorem.  We shall briefly summarize the proof of the
post-prior equivalence in relativistic reactions of composite
particles using Dirac's constraint dynamics \cite{dirac}.  A more
detailed description can be found in Ref.\ [3].

\section{Relativistic 2-Body Bound State Problem}

Before we examine the reaction of composite two-body systems, it is
important to discuss the structure of the bound states.  Much progress
has been made in the study of the relativistic two-body bound state
problem using Dirac's relativistic constraint dynamics
\cite{dirac,tod,cra82,saz85}.  This constraint formalism represents a
simplified, yet complete, resummation of the Bethe-Salpeter
equation. It is similar to many other three-dimensional quasipotential
truncations, but has a number of advantages over many of those
approaches.  We briefly review here the results for bound states of
spinless particles. One assumes a generalized mass shell constraint
for each particle
\begin{equation}
{\cal H}_{i}|\psi \rangle=0{~~~~~~{\rm for~~~~~~~~~}}i=1,2  
\nonumber
\end{equation}
where 
\begin{equation}
{\cal H}_{i}=p_{i}^{2}-m_{i}^{2}-\Phi _{i},
\end{equation}
and $\Phi _{1}$ and $\Phi _{2}$ are two-body interactions dependent on
coordinate $ x_{12}$. One constructs the total Hamiltonian ${\cal H}$
\begin{equation}
{\cal H=\lambda }_{1}{\cal H}_{1}+\lambda _{2}{\cal H}_{2},  \label{2.5}
\end{equation}
where $\{ \lambda _{i}\}$ are Lagrange multipliers.  In order that the
constraints be compatible, we must have
\begin{equation}  \label{eq:com}
\lbrack {\cal H}_{1},{\cal H}_{2}]|\psi \rangle=0.  \label{3}
\end{equation}
The simplest way to satisfy the above condition is to take 
\begin{equation}  \label{eq:res}
\Phi _{1}=\Phi _{2}=\Phi (x_{\perp }),  \label{8}
\end{equation}
where 
\begin{equation}
x_{\nu \perp }=x_{12}^{\mu }(\eta _{\mu \nu }-P_{\mu }P_{\nu }/P^{2}),
\label{9}
\end{equation}
and $P$ is the total momentum $P=p_{1}+p_{2}$.  

The equation of motion, ${\cal H}|\psi \rangle$$=$$0$, describes both
the center-of-mass and the relative motion.  To separate the
center-of-mass and the relative motion, we introduce the relative
momentum $q$
\begin{equation}
p_{1}={\frac{p_{1}\cdot P}{P^{2}}}P+q,  \label{eq:213}
\end{equation}
\begin{equation}
p_{2}={\frac{p_{2}\cdot P}{P^{2}}}P-q.  \label{eq:214}
\end{equation}
We then obtain $
{\cal H}$ in terms of $P$ and $q$: 
\begin{eqnarray}
{\cal H}|\psi \rangle=
(\lambda _{1}+\lambda _{2}) \lbrack q^{2}-\Phi
(x_{\perp })  + b^{2}(P^{2};m_{1}^{2},m_{2}^{2}) \rbrack |\psi
\rangle=0, \nonumber
\end{eqnarray}
where 
$$
b^{2}(P^{2},m_{1}^{2},m_{2}^{2})
=\frac{1}{4P^{2}}
[P^{4}-2P^{2}(m_{1}^{2}+m_{2}^{2})+(m_{1}^{2}-m_{2}^{2})^{2}].
$$ 
The equation of motion, ${\cal H}|\psi \rangle$$=$$0$, can be solved
by introducing the bound state mass $M$ to separate it into the
following two equations for center-of-mass motion and relative
motion
\begin{equation}
\left\{ P^{2}-M^{2}\right\} |\psi \rangle=0,  \label{pm}
\end{equation}
\begin{equation}
(\lambda _{1}+\lambda _{2})\left\{ q^{2}-\Phi (x_{\perp
})+b^{2}(M^{2},m_{1}^{2},m_{2}^{2})\right\} |\psi \rangle=0.  \label{eig}
\end{equation}
The equation for relative motion is independent of $\lambda_i$.  We go
to the center-of-momentum system where $q=q_{\perp }=(0,{\bf q)}$ and
$x_{\perp }=(0, {\bf r)}$.  It is convenient to choose $\lambda
_{i}=1/2m_{i}$ so that the equation for relative motion, Eq.\ (11),
matches the Schr\"{o}dinger equation term by term:

\begin{equation}
\left\{ {\frac{{\bf q}^{2}}{2\mu }}+{\frac{\Phi ({\bf r})}{2\mu }}
\right \} |\psi\rangle=
{\frac{ b^{2}}{2\mu }} |\psi \rangle
= E |\psi \rangle,  \label{eq:eig}
\end{equation}
where $\mu=m_1m_2/(m_1+m_2)  $.
From the eigenvalue of this
Schr\"{o}dinger equation and 
$b^{2}(M^{2},m_{1}^{2},m_{2}^{2})=2\mu E$, we can
obtain the bound state mass $M$ given by 
\begin{equation}
M=\sqrt{2\mu E+m_{1}^{2}}+\sqrt{2\mu E+m_{2}^{2}}.  \label{eq:ME}
\end{equation}

\section{ Scattering of Two Composite Particles}

For a system of two composite particles with a total of $4$
constituents, we can similarly specify a generalized mass shell
constraint for each constituent:
\begin{equation}
{\cal H}_{i}|\psi \rangle=0{~~~~~~{\rm for~~~~~}}i=1,..,4  \label{III1}
\end{equation}
where \cite{saz}
\begin{equation}
{\cal H}_{i}=p_{i}^{2}-m_{i}^{2}- \sum_{j\ne i}^{4}\Phi _{ij}.
\end{equation}
We construct the total relativistic Hamiltonian
as 
\begin{equation}
{\cal H=}\sum_{i=1}^{4}\lambda _{i}{\cal H}_{i},  \label{4}
\end{equation}
and choose $\lambda _{i}=1/2m_{i}$.  The total Hamiltonian is then
\begin{eqnarray}
\label{eq:17}
{\cal H}=\sum_{i=1}^{4}{ p_{i}^{2}-m_{i}^{2} \over 2 m_i}-\sum_{i=1}^{4}
\sum_{j\ne i}^{4}V_{ij},  \label{11}
\end{eqnarray}
where $\mu _{ij}=m_{i}m_{j}/(m_{i}+m_{j})$ and $V_{ij}=\Phi
_{ij}/2\mu _{ij}$.  

The quadratic form of the momentum operators $p_{i}^2$ in this
$4$-body Hamiltonian simplifies the separation the center-of-mass
momentum and the relative momentum for any pair of particles.  One can
therefore use the two-body Hamiltonians to generate basis states.
Specifically, we solve for the bound states $\psi_{ij}$ of mass
$M_{ij}$ for particles $i$ and $j$ interacting with the interaction
$V_{ij}$,
\begin{equation}
\label{eq:18}
{\cal H}_{ij}|\psi _{ij} \rangle=\left \{ {p_{i}^{2}-m_{i}^{2} \over 2
m_i}
+{ p_{j}^{2}-m_{j}^{2} \over 2 m_j} -V_{ij}\right \} |\psi _{ij}\rangle=0.
\end{equation}
To study the relativistic constituent-interchange process of Eq.\
(\ref{eq:1}) as a generalization of the non-relativistic case
investigated by Barnes and Swanson \cite{Bar92}, we can partition the
total Hamiltonian into an unperturbed part ${\cal H}_{0}$ and a
residual interaction $V_{I}$ in two different ways. In the ``prior''
form,  this partition is 
\begin{eqnarray}
{\cal H}={\cal H}_{12}+{\cal H}_{34}-V_{13}-V_{14}-V_{23}-V_{24}, 
\end{eqnarray}

\vspace*{-1cm}
\begin{eqnarray}
{\cal H}_{0}({\rm prior})={\cal H}_{12}+{\cal H}_{34}, 
\end{eqnarray}

\vspace*{-1cm}
\begin{eqnarray}
V_{I}({\rm prior})=-V_{13}-V_{14}-V_{23}-V_{24}. 
\end{eqnarray}
The corresponding ``prior'' reaction matrix element is 
\begin{eqnarray}
\label{hij}
& &2\pi \delta ^{(4)}(P_{A}+P_{B}-P_{C}-P_{D})h_{fi}({\rm
prior})\nonumber \\
&=&-\langle \psi
_{14}\psi _{23}|V_{13}+V_{14}+V_{23}+V_{24}|\psi _{12}\psi _{34}\rangle .
\end{eqnarray}
If we represent the interaction $V_{ij}$ n diagrammatic form
\cite{Bar92} by a curly line, the four terms in the above matrix
element are represented by the four diagrams in Fig.\ 1. The
interaction takes place before the interchange of constituents.

\vspace*{4.5cm} \epsfxsize=300pt \includegraphics{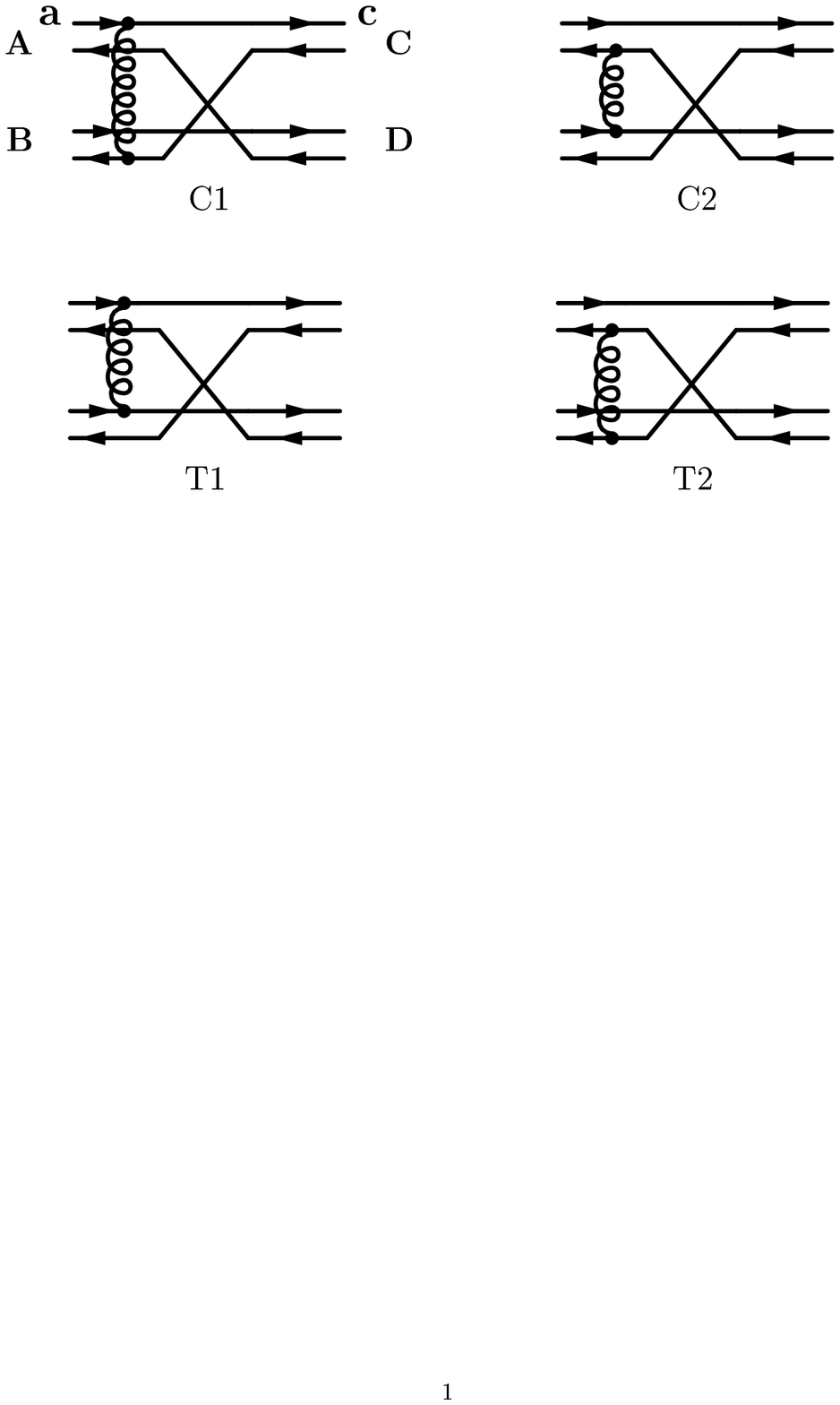} 
\vspace*{0.4cm}\hspace*{-0.7cm} 
\begin{minipage}[t]{14cm}
\noindent {\bf Fig.\ 1}.  {`Prior' diagrams for the reaction A+B
$\to$ C+D.}
\end{minipage}
\vskip4truemm \noindent On the other hand, in the ``post'' form of
partitioning the total Hamiltonian, we have
\begin{eqnarray}
{\cal H}={\cal H}_{14}+{\cal H}_{32}-V_{13}-V_{12}-V_{43}-V_{42}, 
\end{eqnarray}

\vspace*{-1cm}
\begin{eqnarray}
{\cal H}_{0}({\rm post})={\cal H}_{14}+{\cal H}_{32}, 
\end{eqnarray}
\vspace*{-1cm}
\begin{eqnarray}
V_{I}({\rm post})=-V_{13}-V_{12}-V_{43}-V_{42}.
\end{eqnarray}
The ``post'' reaction matrix element is 
\begin{eqnarray}
& &2\pi \delta ^{4}(P_{A}+P_{B}-P_{C}-P_{D})h_{fi}({\rm post})
\nonumber\\
&=&-\langle \psi
_{14}\psi _{23}|V_{12}+V_{13}+V_{42}+V_{43}|\psi _{12}\psi _{34}\rangle .
\end{eqnarray}
The four terms in the post matrix element are represented by the four
diagrams of Fig.\ 2. The interaction takes place after the interchange
of constituents.

\vspace*{4.5cm} \epsfxsize=300pt \includegraphics{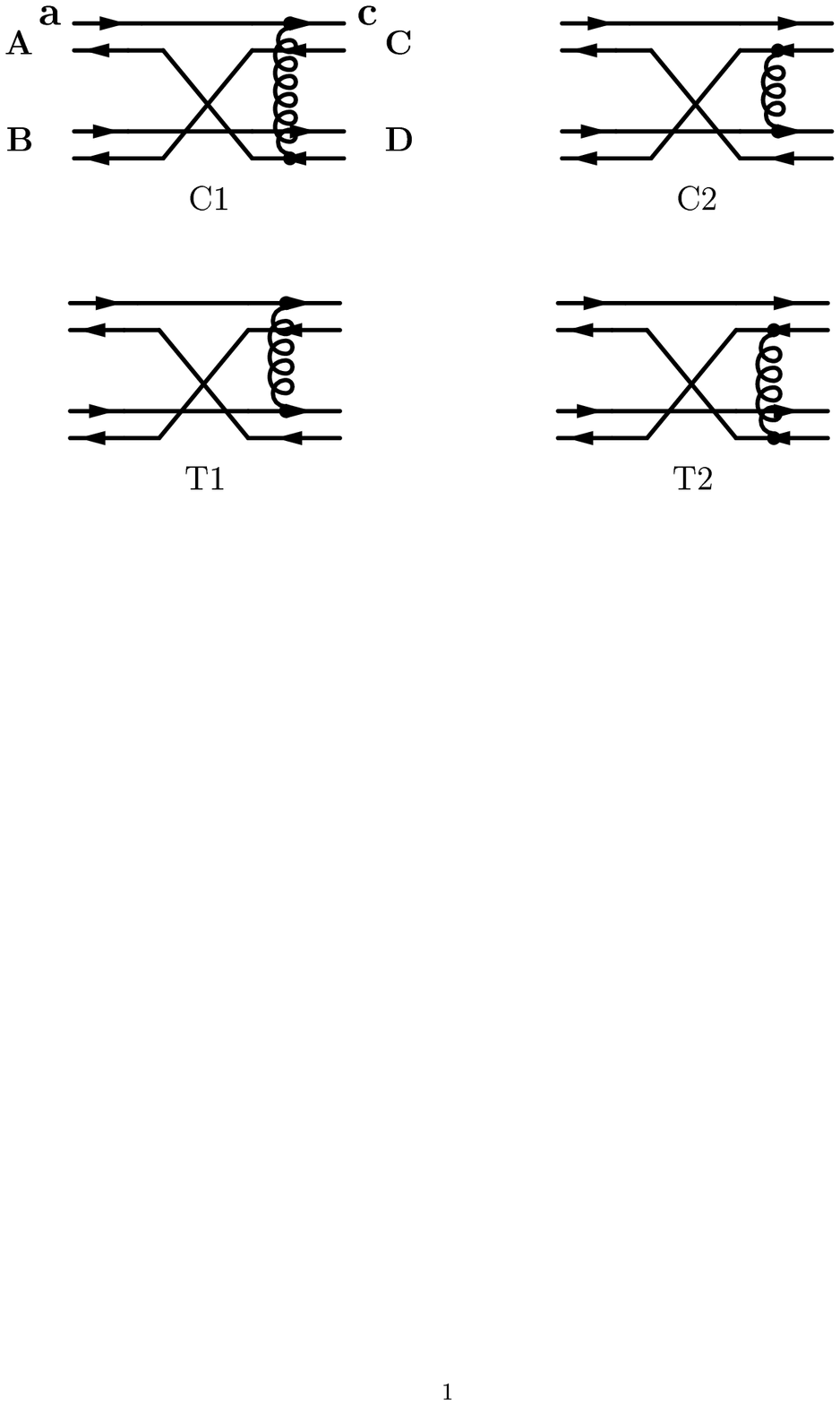} 
\vspace*{0.4cm}\hspace*{-0.7cm} 
\begin{minipage}[t]{14cm}
\noindent {\bf Fig.\ 2}.  
{`Post' diagrams for the reaction $A$+$B$ $\to$ $C$+$D$ .}
\end{minipage}
\vskip 4truemm \noindent

If we start with the prior expression for the matrix
element, since ${\cal H}_{12}|\psi _{12}\rangle=0$ and ${\cal
H}_{34}|\psi _{34}\rangle=0$, it follows that 
\begin{eqnarray}
& &\langle \psi _{14}\psi _{23}|V_{13}+V_{14}+V_{23}+V_{24}|\psi _{12}\psi
_{34}\rangle 
\nonumber\\
&=&\langle \psi _{14}\psi _{23}|-{\cal H}_{12}-{\cal H}%
_{34}+V_{13}+V_{14}+V_{23}+V_{24}|\psi _{12}\psi _{34}\rangle  \nonumber \\
&=&\langle \psi _{14}\psi _{23}|-{\cal H}_{14}-{\cal H}%
_{23}+V_{12}+V_{13}+V_{42}+V_{43}|\psi _{12}\psi _{34}\rangle,
\nonumber
\end{eqnarray}
where we have used Eqs.\ (\ref{eq:17}) and (\ref{eq:18}) for 
the Hamiltonian of the system. Because ${\cal H}_{14}|\psi
_{14}\rangle=0$ and ${\cal H} _{23}|\psi _{23}\rangle=0$, the above
equation leads to 
\begin{eqnarray}
& &\langle \psi _{14}\psi _{23}|V_{13}+V_{14}+V_{23}+V_{24}|\psi _{12}\psi
_{34}\rangle \nonumber\\
&=&\langle \psi _{14}\psi _{23}|V_{12}+V_{13}+V_{42}+V_{43}|\psi
_{12}\psi _{34}\rangle ,
\end{eqnarray}
which is the relativistic generalization of the
post-prior equivalence of the reaction matrix elements.

Just as in non-relativistic reaction theory \cite{Sch68}, the
equivalence is possible if and only if the interaction which induces
the reaction is the same as the interaction that generates the
composite bound states, and the reaction matrix element is evaluated
using these bound state wave functions.  The post-prior equivalence
guarantees a unique result for the reaction cross section in the
first-Born approximation.

\section{Conclusion and Summary}

The formulation of the relativistic two-body bound state problem in
constraint dynamics allows a simple separation of center-of-mass and
relative motion. The equation for relative motion can be cast in the
form of a non-relativistic Schr\"{o}dinger equation in the
center-of-momentum system.  The two-body bound state mass in the
relativistic case is related to the eigenvalue of the Schr\"{o}dinger
equation by a simple algebraic equation, Eq.\ (\ref{eq:ME}).

The relativistic constraint dynamics can be generalized for a
many-particle system.  Using the two-body solution as basis states for
multi-particle dynamics, the $N$-body Hamiltonian can be separated
into an unperturbed Hamiltonian and a residual interaction.  The study
of the dynamics involves the evaluation of the reaction matrix
elements of the residual interaction using the wave functions of the
composite particles.

In rearrangement reactions, because of the the freedom of partitioning
the total Hamiltonian into different unperturbed and interaction
parts, the evaluation of the reaction matrix elements can be carried
out either in the post or prior forms.  We show explicitly the
equality of the post and prior reaction matrix elements for
relativistic reaction of composite particles which guarantees a unique
perturbation expansion in relativistic reaction of composite
particles.

\section*{Acknowledgments}

The authors would like to thank Profs. T. Barnes and
E. S. Swanson for stimulating discussions.  This work was supported by
the Department of Energy under contract DE-AC05-00OR22725 managed by
UT-Battelle, LLC.

\section*{References}


\begin{thebibliography}{99}

\bibitem{Sch68}  L. I. Schiff, {\it Quantum Mechanics} (McGraw-Hill, New
York, 1968), pp. 384-387.

\bibitem{Won00}
C. Y. Wong, E. S. Swanson, and T. Barnes,
Phys. Rev. {\bf C62}, 045201 (2000).


\bibitem{Won01}  C. Y. Wong and H. W. Crater,
Phys. Rev. {\bf C63}, 044907 (2001).


\bibitem{dirac}  P. A. M. Dirac, Canad. J. Math. {\bf 2}, 129 (1950); Proc.
Roy. Soc. Sect. A {\bf 246}, 326 (1958); {\bf Lectures on Quantum Mechanics}
(Yeshiva University, New York, 1964).

\bibitem{tod} I. T. Todorov, Phys. Rev. {\bf D3}, 2351 (1971).

\bibitem{cra82}  P. Van Alstine and H. W. Crater, J. Math. Phys. {\bf 23},
1997 (1982); H. W. Crater and P. Van Alstine, Ann. Phys. (N.Y.) {\bf 148},
57 (1983);  H. W. Crater, R. Becker, C. Y. Wong, and P. Van
Alstine, Phy. Rev. {\bf D46}, 5117 (1992);
H. W. Crater, C. W. Wong, and C. Y. Wong, Int. J. Mod.
Phys. E {\bf 5}, 589 (1996).

\bibitem{saz85} H. Sazdjian, Phys. Lett. {\bf 156B}, 381 (1985);
H. Jollouli and H. Sazdjian, Ann. Phys. {\bf 253}, 376 (1997).

\bibitem{saz}  For a discussion of the compatibility conditions for the
constraints in this case see Refs.\ [7] and [3].

\bibitem{Bar92}  T. Barnes and E. S. Swanson, Phys. Rev. {\bf D46}, 131
(1992);
T. Barnes and E. S. Swanson, Phys. Rev. {\bf C49}, 1166 (1994);
T. Barnes, E. S. Swanson and J. Weinstein, Phys. Rev. {\bf D46}, 4868
(1992).


\end{thebibliography}
\end{document}